# Label-free detection of exosomes from different cellular sources based on surface-enhanced Raman spectroscopy combined with machine learning models

Yang Li[a,b,c,1,*], Xiaoming Lyu[b,1], Kuo Zhan[a], Haoyu Ji[d], Lei Qin[b], Jian-An Huang[a]

a. Research Unit of Health Sciences and Technology (HST), Faculty of Medicine University of Oulu, Finland

b. Research Center for Innovative Technology of Pharmaceutical Analysis, College of Pharmacy，Harbin Medical University，Heilongjiang 150081, PR China

c. National Key Laboratory of Frigid Zone Cardiovascular Diseases (NKLFZCD), College of Pharmacy, Harbin Medical University, Heilongjiang 150081, PR China;

d. Department of Pharmacy at The Second Affiliated Hospital, and Department of Pharmacology at College of Pharmacy (The Key Laboratory of Cardiovascular Medicine Research, Ministry of Education), Harbin Medical University, Harbin 150081, PR China

* Corresponding authors:
Email address: liy@hrbmu.edu.cn

[1] These authors contributed equally to this work.

# ABSTRACT

Exosomes are significant facilitators of inter-cellular communication that can unveil cell-cell interactions, signaling pathways, regulatory mechanisms and disease diagnostics. Nonetheless, current analysis required large amount of data for exosome identification that it hampers efficient and timely mechanism study and diagnostics. Here, we used a machine-learning assisted Surface-enhanced Raman spectroscopy (SERS) method to detect exosomes derived from six distinct cell lines (HepG2, Hela, 143B, LO-2, BMSC, and H8) with small amount of data. By employing sodium borohydride-reduced silver nanoparticles and sodium borohydride solution as an aggregating agent, 100 SERS spectra of the each types of exosomes were collected and then subjected to multivariate and machine learning analysis. By integrating Principal Component Analysis with Support Vector Machine (PCA-SVM) models, our analysis achieved a high accuracy rate of 94.4% in predicting exosomes originating from various cellular sources. In comparison to other machine learning analysis, our method used small amount of SERS data to allow a simple and rapid exosome detection, which enables a timely subsequent study of cell-cell interactions, communication mechanisms, and disease mechanisms in life sciences.

# 1. Introduction

Exosomes are extracellular vesicles that are actively released by living cells, and their diameters usually range from 30-150 nm. Exosomes contain nucleic acids, proteins, and lipids, and their function as an important communication medium between cells has attracted much attention(Hessvik and Llorente 2018; Špilak et al. 2021; Staubach et al. 2021). Initially, exosomes were thought to be carriers of cellular wastes and were widely found in human fluids, such as saliva, sweat, blood, and urine(Jara-Acevedo et al. 2019). However, more and more studies have demonstrated that the content of specific components of exosomes can reflect the pathophysiological status of parental cells, and the protein, nucleic acids, and lipid content vary with cellular composition and function(Teng and Fussenegger 2021). Thus, exosomes play an important role in intercellular communication, tumour development, and metastasis(Hu et al. 2020; Meldolesi 2018; Regev-Rudzki et al. 2013; Vyas and Dhawan 2017). Liquid biopsy methods for exosomes reduce patient harm and lower testing costs compared with traditional invasive histopathology biopsies. Therefore, exosomes have been widely studied as biomarkers for tumours(Han et al. 2022; Li et al. 2022a; Li et al. 2022b; Pan et al. 2021; Zhou et al. 2020). Traditional methods for exosome detection include transmission electron microscopy (TEM), nanoparticle tracking analysis (NTA), protein immunoblotting (Western blot), and enzyme-linked immunosorbent assay (ELISA) methods(Shao et al. 2018; Ueda et al. 2014). However, these methods are time-consuming and cumbersome. In recent years, many new exosomes detection methods have been reported, such as fluorescence-integrated microfluidics(Liu et al. 2018), electrochemical sensors(Cao et al. 2020), and CRISPR/CAS system-assisted detection(Zhao et al. 2020b), which have improved the sensitivity and detection limit of detecting exosomes; however, the complex exosomes pretreatment process and technical operation limit their application.

Surface-enhanced Raman spectroscopy (SERS) has been widely used in the field of analytical detection due to its simple operation, rapid detection, easy sample handling, high sensitivity, and nondestructiveness(Huang et al. 2019; Jiang et al. 2021; Wang et al. 2022; Zhang et al. 2023; Zhao et al. 2020a). In recent years, SERS technology has been widely used in the qualitative and quantitative analysis of exosomes. The detection of exosomes is divided into two methods, labelled and label-free(Shin et al. 2020), and labelled detection is mainly by labelling proteins on exosomes. For example, Zong et al. fabricated anti-CD 63 antibody-modified magnetic nanoparticles and anti-HER 2 antibody-modified Au@Ag nanorods for qualitative and quantitative detection of exosomes(Zong et al. 2016). Pang et al. used anti-PD-Ll antibody-modified Au@Ag@MBA as a SERS tag to label exosomes with PD-Ll for quantitative purposes(Pang et al. 2020). Han et al. used protein biomarkers for the diagnosis of osteosarcoma based on the exosomes. analysis for the diagnosis of osteosarcoma(Han et al. 2022). These studies were highly accurate and reproducible, but the use of tags raised the cost and limited the scope of application. To overcome these problems, researchers have developed label-free exosomes assays, such as Lee et al. who used a silver-film-coated nanobowl platform to capture exosomes secreted by SKOV 3 cells(Lee et al. 2015). Stremersch and colleagues prepared positively charged Au NPs to detect and differentiate exosomes from different cellular sources by coating them with a 4-dimethyl aminopyridine layer(Stremersch et al. 2016). Dong et al. developed a gold-coated $TiO_2$ macroporous inverse opal (MIO) structure to capture and analyze exosomes from the plasma of cancer patients(Dong et al. 2020). These researchers designed novel SERS substrates to enable label-free detection of exosomes, which broadens the range of applications compared to labelled methods. However, the synthesis process of these substrates is relatively complex and costly. Therefore, the development of a time-saving and convenient exosome detection method is of great significance for exosome research.

Machine Learning, the branch of Artificial Intelligence, can learn and improve from data sets. With algorithms such as Partial Least Squares Regression (PLS), Linear Discriminant Analysis (LDA), Random Forests (RF), Support Vector Machines (SVM), and Convolutional Neural Networks (CNN), Machine Learning can analyze large amounts of data, spot patterns and trends, and make predictions or make decisions based on these features. Due to the approximation of Raman spectra as multivariate data, machine learning has been increasingly integrated with surface-enhanced Raman spectroscopy in recent years(Bi et al. 2023; Ju et al. 2023; Lin et al. 2022; Lussier et al. 2020), including Zhe Zhang et al. who combined linear discriminant analysis (LDA) to achieve classification and identification of adenoviruses(Zhang et al. 2024), Janina Kneipp et al. who used a random forest-based approach for analyzing lipid signals that contribute weakly to cellular SERS spectra(Živanović et al. 2019), and Hyunku Shin et al. used a Convolutional Neural Networks (CNN) model to analyze the surface-enhanced Raman spectral features of exosomes for the simultaneous detection of six early cancers(Shin et al. 2023). Due to the excellent performance of the SVM algorithm in machine learning in handling small sample data sets, the decision boundary is more precise, overfitting can be effectively avoided, and the model results are more accessible to interpret. It has been more widely used(Dies et al. 2018; Huang et al. 2023; Li and Chin 2021; Wang et al. 2015).

In this study, we label-free detected exosomes from different cellular sources and combined them with the PCA-SVM algorithm for differentiation and prediction (Fig. 1). We first prepared silver nanoparticles by reducing silver nitrate with sodium borohydride as a substrate for Surface-enhanced Raman spectroscopy (SERS) and used sodium borohydride solution as an aggregation agent for detecting exosomes signals(Wang et al. 2023). We successfully obtained SERS spectra of exosomes from three cancer cells and their corresponding normal cells. To further investigate the effectiveness of this method in exosomes detection from different cell sources, We applied principal component analysis (PCA) for dimensionality reduction of SERS data and combined it with support vector machine (SVM) algorithm in machine learning for analysis and prediction. The experimental results show that this method is able to obtain the signals of exosomes from different sources without labelling, easily and quickly, and classify and identify them.

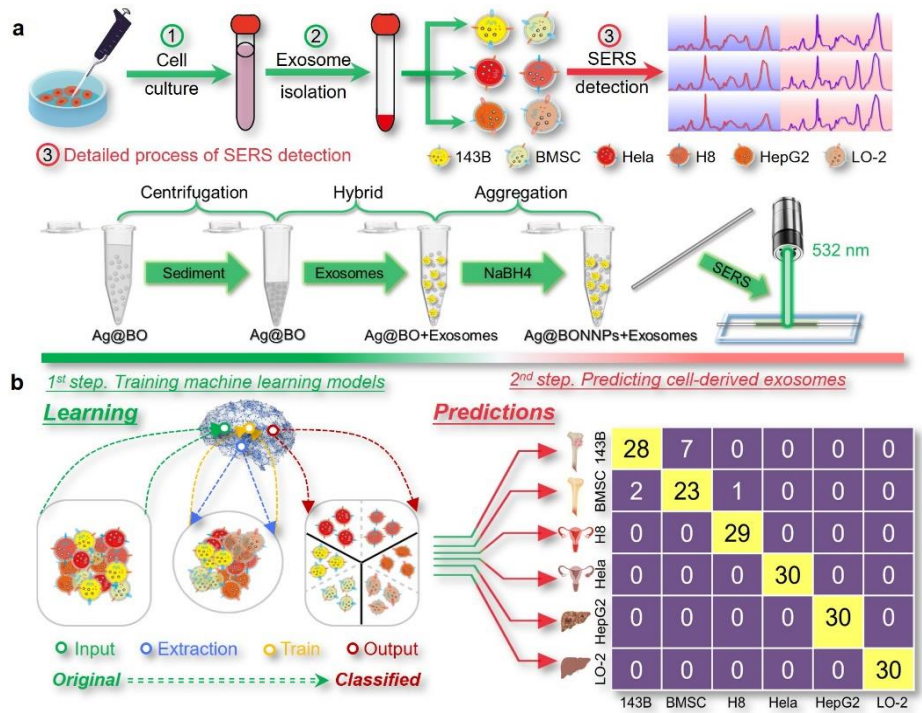

*Figure 1: Schematic diagram of exosome SERS detection and SVM machine model differentiation for different cellular sources. a. Extraction of cell-derived exosomes and SERS detection process. Exosomes were extracted from cells after starvation culture, and Ag@ BONPs and NaBH4 solutions were added sequentially, mixed, and assayed to obtain SERS profiles. b. SVM machine learning model training for SERS identification and differentiation of cell-derived exosomes. In the first step, 70% of the data is input as a training set, and the support vector machine model is trained after extracting the main features. In the second step, 30% of the data is used as a test set for prediction to get a confusion matrix. Cartoons were created with BioRender.com.*

## 2. Experimental methods

### 2.1 Cell culture

DMEM high glucose medium was purchased from Gibco, F12 medium was purchased from Sigma-Aldrich, and all cell lines were purchased from American Model Culture Collection (ATCC, China), of which Hela, H8, HepG2, LO-2, and 143B cells were cultured in 500 ml of DMEM high glucose medium with 10% fetal bovine serum (FBS) and 1 mM penicillin/streptomycin. (FBS) and 1 mM penicillin/streptomycin. BMSC cells were cultured using 500 ml F12 medium to which 10% fetal bovine serum (FBS) and 1 mM penicillin/streptomycin were added. All cells were grown at 37°C and 5% $CO_2$, the medium was changed every 2-3 days, and the cells were passaged after reaching 90% cell fusion. When EV was collected, the passaged cells were cultured until confluent with the above growth medium.

The cells were then washed with PBS (3 times) and starved in medium prepared in the presence of fetal bovine serum-free for 48 h. After 48 h, the cell supernatant was taken for exosomes isolation.

## 2.2 Exosomes isolation and characterization

After starvation culture, the cell supernatant was obtained, and the cellular exosomes were obtained by ultracentrifugation, the supernatant was first centrifuged at 300×g for 10 min to remove cells, 3000×g for 20 min to remove dead cells, and 10,000×g for 30 min to remove cellular debris, and then centrifuged at 120,000×g for 70 min to obtain the exosomes crude, and then resuspended in PBS, and then again 120,000×g PBS resuspension and centrifugation again at 120000×g for 70 minutes to get the purer exosomes. The exosomes obtained were subjected to SERS on the same day to obtain the exosomes fingerprints, and the remaining exosomes samples were stored in the refrigerator at -80°C The exosomes of 143B, Hela, and BMSC cells were selected for TEM, NTA characterization. Transmission electron microscopy (TEM) revealed teacup-shaped exosomes, and nanoparticle tracer (NTA) results showed that the diameters of the samples were concentrated around 100~150 nm. Western blotting was performed to compare the exosomes samples of 143B, Hela, and BMSC cells with the exosomes surface proteins, CD9, and CD81 labelling, and the negative indicator, Calnexin protein. Protein electrophoresis comparison, TEM, NTA, and Western Blot results showed that we did obtain the exosomes of the above cells.

## 2.3 SERS substrate preparation and detection

Enhanced substrates were synthesized by the method we used previously, where 5 mL of silver nitrate solution (6.6 mg/ mL) was added to 495 mL of sodium borohydride solution (0.133 mg/ mL), stirred for 18 min, centrifuged at 5500 rpm, 25°C for 20 min, and the supernatant was removed, and 10 μL of the silver nitrate lysate, mixed with 10 μL of the cellular exosomes and 5 μL of sodium borohydride (0.05 M, pH=10) were mixed. The solution was stirred homogeneously and then subjected to SERS detection using a WITec Alpha 300 R (Ulm, Germany) instrument. The laser wavelength was 532 nm, the scan time was 30 s, the energy was 20 mW, and the cumulative number of times per test was one.

## 2.4 Data Processing and Machine Learning

We obtained 100 SERS spectra of each of these six cellular exosomes, and the spectra were baseline corrected by LabSpec6 software, and the (600-1800 $cm^{-1}$) data range was selected for smoothing, and then the spectral data were normalized by Origin software to obtain the processed data.

Principal Component Analysis (PCA) is a linear feature extraction technique, which projects the data into the low dimensional space by linear mapping, by doing so it can ensure that the original data has the highest variance in the low dimensional space. It does this by calculating the eigenvectors in the covariance matrix of its features. The eigenvectors corresponding to the largest eigenvalues (principal components) - are used to reconstruct new data and it is ensured that these data have the maximum variance in the direction of that eigenvector.

Hierarchical Cluster Analysis (HCA) is a cluster analysis method that aims to establish a hierarchical structure of clusters. It allows similar items to be grouped into clusters based on their characteristics and is an unsupervised learning algorithm. In hierarchical cluster analysis, data points are initially treated as individual clusters, which are then sequentially merged or divided based on their similarity until a single cluster containing all data points is formed. This process produces a hierarchical tree structure, called a dendrogram, which visually represents the relationships between clusters.

Support Vector Machine (SVM) are a class of generalized linear classifiers that binary classify data in a supervised learning fashion, and are linear classifiers defined to maximize the interval over the feature space, while the kernel trick also included in SVM makes him essentially nonlinear classifiers. Therefore we choose SVM as the model for machine learning.

The processed spectral data was imported into RStudio and PCA downscaling was performed on the data using the prcomp function package in RStudio. Support Vector Machine (SVM) classifiers are built using the dimensionality reduced data features PC1, PC2, PC3 and parameter search results to get the best classifier parameters, after building the best classifier model 70% of the data is used as the training set and 30% of the data is used as the test set to test the accuracy of the model.

## 3. Results and discussion

### 3.1 Isolation and characterization of exosomes.

Ultracentrifugation is the gold standard for the isolation of exosomes(Royo et al. 2020), including differential centrifugation and density gradient centrifugation, in which differential centrifugation does not require the addition of additional reagents and does not interfere with the process of label-free detection of exosomes; therefore, we used differential centrifugation to purify and isolate exosomes from cell lines after starvation culture, and the specific experimental procedure is shown in Fig. 2.a. The obtained exosomes were characterized using conventional experiments such as TEM, NTA, and Western blot. TEM results showed (Figure 2. b) that the exosomes we obtained by differential centrifugation were between 100~150 nm in size and had no prominent protein aggregates. The particle size obtained by NTA (Figure 2. c) further confirmed the results (see Supplementary Fig. 1 for more information).

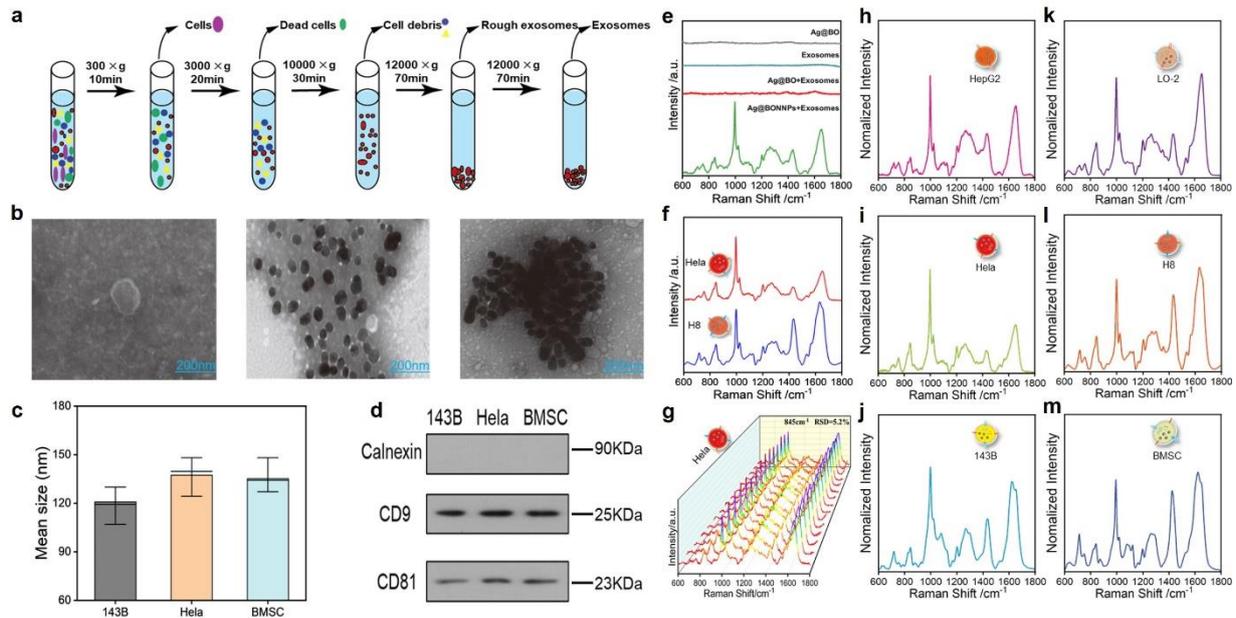

*Figure 2: Isolation, characterization, and SERS mapping of cell-derived exosomes. a. Isolation and extraction process of cell-derived exosomes. b. TEM of cell-derived exosomes, TEM of exosomes with Ag@BO NPs, TEM of exosomes with Ag@BONNPS, scale bar: 200 nm. c. NTA particle size results of the three cell-derived exosomes of 143B, Hela, BMSC. d. 143B, Hela, BMSC WB results of three cell-derived exosomes showing protein blotting of exosome marker expression. e. SERS profiles of Hela cell-derived exosomes under different systems. f. SERS profiles of Hela cell and H8 cell-derived exosomes. g. Fifteen sets of randomized Hela cell-derived exosome profiles were obtained using the Ag@ BONPs method. The RSD was 5.2% at a peak intensity of 845 cm-1. h-m. SERS profiles of exosomes derived from six different cell sources.*

Common exosomes markers (CD9 and CD63) were selected as positive control proteins, and Calnexin protein was used as a negative control protein for Western blot, and the protein blot bands (Fig. 2.d) indicated that we successfully obtained cell-derived exosomes from cell culture fluid.

### 3.2 SERS detection of exosomes.

The SERS-enhanced substrate used for detecting cell-derived exosomes was obtained using the method we have used before(Zhang et al. 2022). Ag@BO nanoparticles were obtained by reducing silver nitrate solution using sodium borohydride and centrifuging the solution. The specific detection process is shown in Fig. 1.a. The dynamic light spectroscopy results of the mixed solution of each part are shown in Supplementary Fig. 2, where there was no SERS signal for both Ag@BO nanoparticles and exosomes. We mixed Ag@BO nanoparticles with cellular exosomes, and still no signal was detected, as shown in Fig. 2.b. Ag@BO nanoparticles were not tightly bound together after mixing with exosomes, and when we added sodium borohydride solution as an aggregating agent, Ag@BO nanoparticles were bound to the amino group on the surface of the exosomes to form silver-ammonia covalent bonds, and Ag@BO

nanoparticles were adsorbed on the exosomes' surface, like a " cleanser ", we were able to get more material information about the exosome, and therefore we were able to get the signals of the cell-derived exosome(Fig.2.e). The 15 sets of SERS profiles of the exosomes of Hela cells detected at different times had an RSD of 5.2% at the peak position of 845 cm-1 with good reproducibility (Fig. 2.g). Subsequently, we examined exosomes from H8 cells and found that their SERS signals had similar and distinguishing features from those of Hela cell-derived exosomes(Fig. 2.f), indicating the excellent specificity of our method. We then performed SERS spectroscopy using the same method on exosomes of HepG2, Hela, 143B, LO-2, BMSC, and H8 cell origin. One hundred spectra were acquired from each cell-derived exosomes. These signals were preprocessed, including baseline correction, denoising, and min-max normalization, and peak attribution was performed (see Table 1 of the Supplementary Information for details); the average spectra obtained from the processing of 100 spectral data for each cell are shown in Figure 2.h-m. Since these cell-derived exosomes have the same composition and differ only slightly in content, resulting in slight differences in their SERS spectra, it is difficult for the human eye to observe such differences. Therefore, we used thermogram、principal component analysis, and cluster analysis methods to demonstrate the differences in the main features of spectral data of exosomes from different cellular sources.

### 3.3 Principal component analysis and cluster analysis of SERS profiles of exosomes from cancer and normal cell sources.

We first performed heat map comparison, principal component analysis and hierarchical clustering analysis on the SERS profiles of three different cancer cells and their corresponding normal cells, in which the exosome signals originated from cervical cancer cells (Hela) and cervical epithelial cells (H8) could be completely distinguished from each other, with obvious differences (Fig. 3. c-e). Whereas the difference in exosome signals between hepatocellular carcinoma cells (HepG2) and normal hepatocytes (LO-2) was small, the difference in heatmaps was not obvious, and the results of hierarchical clustering analysis were not satisfactory, but the two could be differentiated by principal component analysis (Fig. 3. f-h), and the heatmap comparisons of exosome signals from osteosarcoma (143B) and bone marrow mesenchymal stromal cells (BMSC) sources were significantly different, and the 95% of the principal component analysis confidence intervals partially overlapped, but hierarchical cluster analysis was able to distinguish the two (Fig. 3. i-k). We further compared the differences in signals between exosomes of cancer cell origin and those of normal cell origin, in which we found a common point that the highest peaks of exosomes of cancer cell origin came from the 1654 cm-1 peak of amide I, i.e., the C=C stretching vibration, whereas the highest peaks of those of normal cell origin came from the symmetric respiratory vibration of phenylalanine. This feature will help to study the molecular mechanisms in cancer cells. To investigate the differences in exosomes from different cellular sources, we used heatmaps to show the spectral signatures of cancer cells from three different tissue sources (143B, Hela and HepG2) (Fig. 4.a-b), and then computed the eigenvectors in the spectral eigen-covariance matrices using unsupervised dimensionality reduction using Principal Component Analysis (PCA) and projected the data into a low-dimensional space by linear mapping (Fig. 4.c.III). The cumulative contribution of PC1 and PC2 reaches 86%, indicating that the first two principal components explain the variability of the spectral data well.

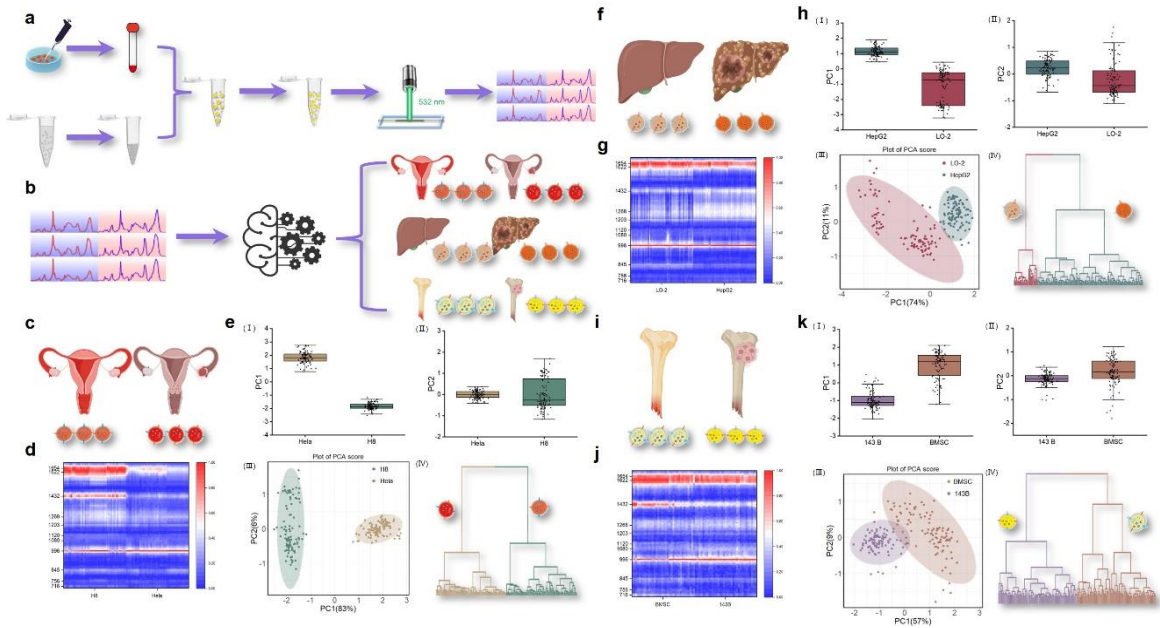

*Figure 3: Detection flow chart, Thermograms, Principal component analysis, and Hierarchical clustering analysis of exosomes from three cancer cells and their corresponding normal cell sources. a. Diagram of the process of detecting cell-derived exosomes. b. Diagram of the process of distinguishing between cancer and normal cell-derived exosomes. c. Schematic diagram of cervical cancer. d. Thermogram comparison of exosomes derived from cervical cancer cells and cervical epithelial cells. e. PCA analysis and HCA analysis of exosomes derived from cervical cancer cells and cervical epithelial cells. (I) and (II) PCA showed significant differences between the exosome profiles of cervical cancer cells and normal cells. (III) PCA analysis of exosomes from cervical cancer cells and normal cells mapped (IV) HCA analysis results confirmed the results of PCA. f. Schematic diagram of hepatocellular carcinoma. g. Thermogram comparison of exosomes derived from hepatocellular carcinoma cells and normal hepatocytes. h. PCA analysis and HCA analysis of exosomes derived from hepatocellular carcinoma cells and normal hepatocytes. (I) and (II) PCA showed significant differences between the exosome profiles of hepatocellular carcinoma cells and normal hepatocytes. (III) PCA analysis of exosomes from hepatocellular carcinoma cells and normal hepatocytes (IV) HCA analysis results confirm the PCA results. i. Schematic diagram of osteosarcoma. j. Comparison of thermograms of exosomes originating from osteosarcoma cells and bone marrow mesenchymal stem cells. k. PCA analysis and HCA analysis of exosomes originating from osteosarcoma cells and bone marrow mesenchymal stem cells. ii. (I) and (II) PCA showed significant differences between osteosarcoma cell and bone marrow mesenchymal stem cell-derived exosomes profiles. (III) PCA analysis maps of exosomes from osteosarcoma cells and bone marrow mesenchymal stromal cells (IV) HCA analysis results confirmed the PCA results. Cartoons were created with BioRender.com.*

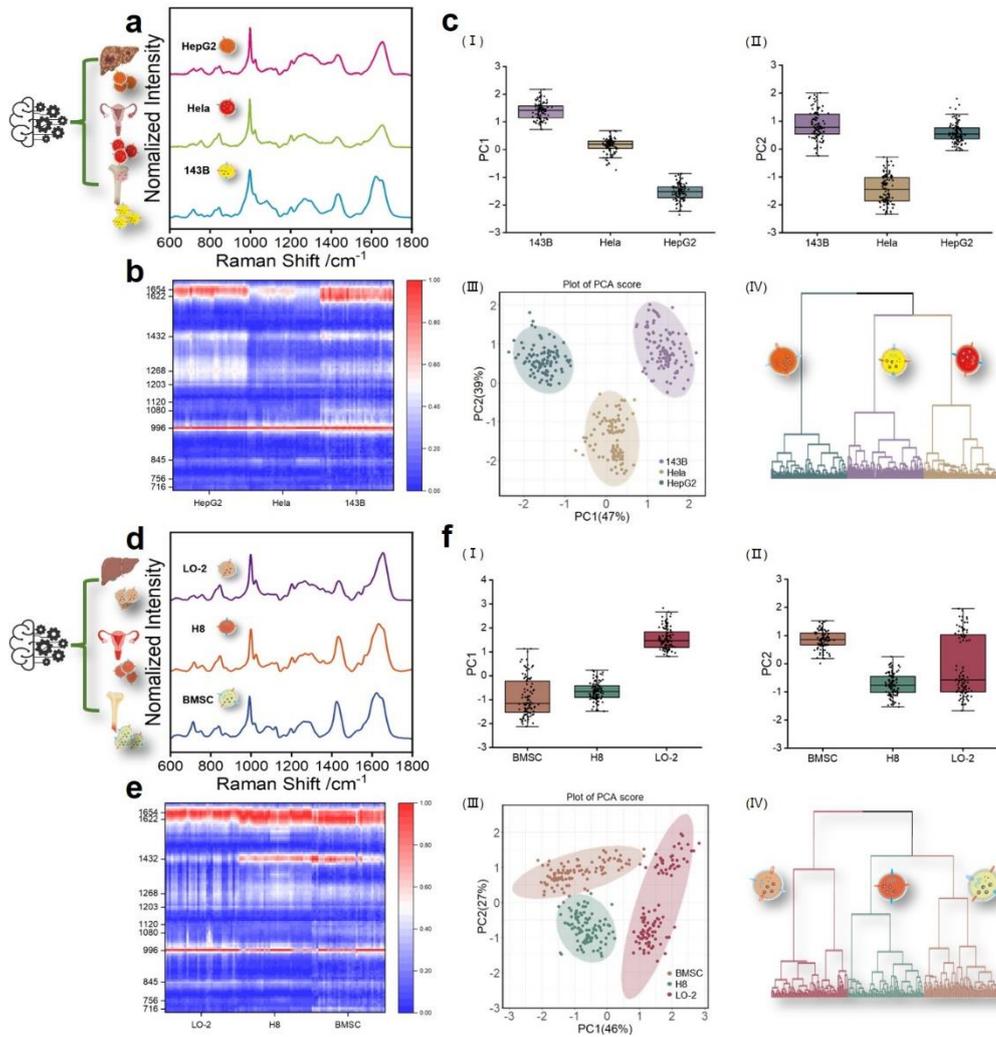

*Figure 4：SERS profiles, thermograms, Principal component analysis, and Hierarchical clustering analysis of three normal cell-derived exosomes and three cancer cell-derived exosomes. a. SERS profiles of exosomes from three cancer cell sources. b. Thermograms of exosomes from three cancer cell sources. c. PCA analysis and HCA analysis of exosomes from three cancer cell sources. (I) and (II) PCA showed significant differences between the exosomes profiles of cancer cells from different tissue sources. (III) PCA analysis of exosomes from different tissue sources of cancer cells mapped (IV) HCA analysis results confirmed the results of PCA. d. SERS profiles of exosomes from three normal cell sources. e. Heat maps of exosomes from three normal cell sources. f. PCA analysis and HCA analysis of exosomes from three normal cell sources. (I) and (II) PCA showed significant differences between the exosomes profiles of normal cells from different tissue sources. (III) PCA analysis of exosomes from different tissue sources of normal cells.(IV) HCA analysis results confirmed the PCA results. Cartoons were created with BioRender.com.*

We were able to find the principal component differences in the SERS spectra of exosomes from cancer cells of different tissue origins by principal component analysis (Fig. 4.c.I-II), and in this way we were able to distinguish them. To further highlight the differences between exosomes from different cellular sources, we chose hierarchical clustering analysis, an unsupervised learning method, to analyse the SERS data. The results of hierarchical clustering analysis based on Euclidean distance were the same as those of principal component analysis, and the SERS profiles of exosomes from the three cancer cell sources were classified into three categories with significant differences (Fig. 4.c.IV). To exclude the specificity of the appearance of differences in exosomes of cancer cell origin, we also demonstrated the exosome signals of the three normal cell origins using heatmaps for principal component analysis and hierarchical clustering analysis (Fig. 4.d-f), and the results of the principal component analysis showed that the cumulative contribution of PC1 and PC2 was 73%, and even though the cumulative contribution was slightly lower than that of the results of the exosomes of cancer cell origin, they could still be distinguished from each other ( Fig. 4.f.I-III), and the results of principal component analysis were further verified by hierarchical clustering analysis (Fig. 4.f.IV). The results showed that there were also significant differences in exosome signals from different normal cellular sources. Therefore, our method obtains the SERS signals of exosomes from different cell sources, and the use of heatmaps can show the signal differences of exosomes more intuitively, using principal component analysis and hierarchical clustering analysis can initially distinguish the exosomal SERS signals of cells from different tissue sources, but it cannot completely and accurately distinguish the exosomal signals of cancer cell sources and normal cell sources, We tried to differentiate the SERS profiles of different cellular exosomes using a support vector machine model, and because the SERS data contains a large number of data features, we used PCA to downsize the data to avoid the appearance of overfitting.so we try to construct a principal component analysis and support Vector Machine (PCA-SVM) combined machine learning model to distinguish exosomal signals from different cell sources and make predictions.

**3.4 Classification and prediction of different cellular exosomes using machine learning models.**

Initially, we found that the conventional 2D PCA could not well distinguish the signals of the six cellular exosomes with most of the overlapping parts (please refer to Fig. 3 in the Supplementary Information). Therefore, we adopted a three-dimensional PCA approach (please refer to Fig. 5.b) and formed a feature dataset by calculating the PC1, PC2 and PC3 eigenvalues of each cellular exosome.We divided this dataset into a 70% training set and a 30% test set and then used the training set to train the SVM model and the test set to validate the performance of the model. The classification interface of the SVM model is shown in supplementary Information Figure 4. Finally, we used the test set to evaluate the model's performance and generated the confusion matrix (Figure 5.c). The detailed results of the confusion matrix can be viewed in Supplementary Information Figure 5. According to the confusion matrix results, our model achieves an accuracy of 94.4%. This indicates that our method can effectively detect the SERS profiles of exosomes of different cellular origins and realize their differentiation and prediction by combining the PCA-SVM machine-learning model.

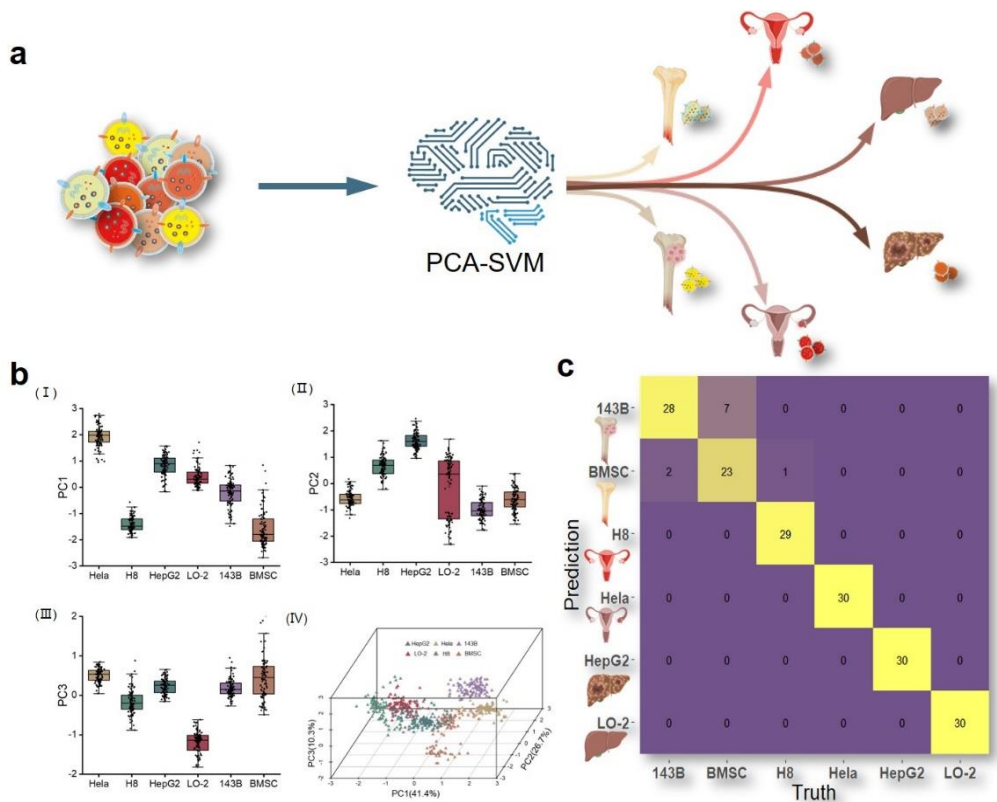

*Figure 5: 3D PCA of exosomes from six different cellular sources and confusion matrix obtained from machine learning results. a. Conceptual diagram of the PCA-SVM machine learning algorithm to distinguish six different cellular sources of exosomes b. (I) and (II) PCA showed significant differences between the six different cellular sources of exosome profiles. (III) PCA analysis plots of exosomes of six different cellular origins. （IV）3D PCA profiles of SERS profiles of exosomes from six cellular sources (PC1=41.4%, PC2=26.7%, PC3=10.3%).c. Prediction results of the three principal components (PC1, PC2, PC3) features of SERS profiles of exosomes of six different cellular origins on 180 sets of untrained test samples based on the obtained support vectors after training using the support vector machine model. Cartoons were created with BioRender.com.*

## 4. Conclusions

In conclusion, we developed a label-free rapid detection of exosomes from different cellular sources. We combined it with a PCA-SVM algorithm to classify and predict exosomes from different cellular sources, and the accuracy of recognizing exosomes from six cellular sources reached 94.4%. Our method is simpler and faster than traditional exosomes detection methods, with better stability, and can also compare the differences in the composition of exosomes from different cellular sources, which can be used to identify exosomes from different cellular sources and to obtain the characteristics of exosomes produced by different cell types. Our research provides a simple and rapid new method for other researchers to study more deeply the mechanism of intercellular communication, the mechanism of disease occurrence, and other life science problems, provides a basis for the implementation of personalized medicine, and provides new ideas and platforms for the development of more effective and targeted exosomes drugs.

**CRediT authorship contribution statement**

Yang Li, Xiaoming Lyu: Experimental part, data processing, machine learning, drawing charts, first draft writing; Haoyu Ji: cell culture, Provide exosome samples and exosome characterisation experiments; Lei Qin: cell culture, Extraction of cellular exosomes; Kuo Zhan: manuscript editing; Jian-An Huang: Conceptualization, Methodology, Writing – review & editing, Supervision, Project administration.

**Declaration of competing interest**

The authors declare that they have no known competing financial interests or personal relationships that could have appeared to influence the work reported in this paper.

**Acknowledgments**

This work was supported by the National Natural Science Foundation for Youth (No. 82202648) and DigiHealth project (project number 326291), a strategic profiling project at the University of Oulu that is supported by the Academy of Finland and the University of Oulu.

**Supporting Information**

Details of TEM,NTA and DLS characterisation of exosomes; 2D principal component analysis plot; Support vector machine classification interface and confusion matrix results; SERS peak attribution for six cell-derived exosomes;